\documentclass[preprint,pdf]{iucr}              

                   \def\href#1{\relax}\let\foo\caption
\ifPDF
  \RequirePackage{hyperref}
  \PassOptionsToPackage{pdftex,bookmarksopen,bookmarksnumbered}{hyperref}
\fi
\let\caption\foo

    \paperprodcode{a000000}      
    \paperref{xx9999}            
    \papertype{FA}               
    \paperlang{english}          
    \journalcode{A}              

    \journalyr{2000}
    \journaliss{1}
    \journalvol{56}
    \journalfirstpage{000}
    \journallastpage{000}
    \journalreceived{0 XXXXXXX 0000}
    \journalaccepted{0 XXXXXXX 0000}
    \journalonline{0 XXXXXXX 0000}

\usepackage[utf8]{inputenc}

\usepackage{hyperref}
\usepackage{bookmark}

\usepackage{bm}
\usepackage{amsmath}
\usepackage{amssymb}
\usepackage{mathtools}

\usepackage{url}

\usepackage{graphicx}

\usepackage{tikz}
\usetikzlibrary{positioning,arrows.meta,shapes}

\usepackage{tabularx}
\newcolumntype{Y}{>{\centering\arraybackslash}X}
\newcolumntype{b}{Y}
\newcolumntype{s}{>{\hsize=.5\hsize}Y}

\newcommand{\relmiddle}[1]{\mathrel{}\middle#1\mathrel{}}
\newcommand{\set}[2]{\left\{ #1 \relmiddle| #2 \right\}}

\newcommand{\software}[1]{\textsc{#1}}
\newcommand{\term}[1]{\emph{#1}}

\begin{document}



\title{
    Algorithms for magnetic symmetry operation search and identification of magnetic space group from magnetic crystal structure
}
\shorttitle{
    Algorithms for deriving MSG information
}


\cauthor[a]{Kohei}{Shinohara}{kshinohara0508@gmail.com}{}
\cauthor[b,c]{Atsushi}{Togo}{atz.togo@gmail.com}{}
\author[a,c,d]{Isao}{Tanaka}{}{}

\aff[a]{Department of Materials Science and Engineering, Kyoto University, \city{Sakyo}, Kyoto 606-8501, \country{Japan}}
\aff[b]{Research and Services Division of Materials Data and Integrated System, National Institute for Materials Science, \city{Tsukuba}, Ibaraki 305-0047, \country{Japan}}
\aff[c]{Center for Elements Strategy Initiative for Structural Materials, Kyoto University, \city{Sakyo}, Kyoto 606-8501, \country{Japan}}
\aff[d]{Nanostructures Research Laboratory, Japan Fine Ceramics Center, \city{Nagoya} 456-8587, \country{Japan}}


\shortauthor{K. Shinohara, A. Togo and I. Tanaka}







\maketitle

\begin{synopsis}
In this paper, algorithms for determining magnetic symmetry operations of magnetic crystal structures, identifying magnetic space-group types from a given magnetic space group (MSG), searching for transformations to a BNS setting, and symmetrizing the magnetic crystal structures on the basis of the determined MSGs are presented.
\end{synopsis}



\begin{abstract}

A crystal symmetry search is crucial for computational crystallography and materials science.
Although algorithms and implementations for the crystal symmetry search have been developed, their extension to magnetic space groups (MSGs) remains limited.
In this paper, algorithms for determining magnetic symmetry operations of magnetic crystal structures, identifying magnetic space-group types of given MSGs, searching for transformations to a BNS setting, and symmetrizing the magnetic crystal structures using the MSGs are presented.
The determination of magnetic symmetry operations is numerically stable and is implemented with minimal modifications from the existing crystal symmetry search.
Magnetic space-group types and transformations to the BNS setting are identified by a two-step approach combining space-group type identification and the use of affine normalizers.
Point coordinates and magnetic moments of the magnetic crystal structures are symmetrized by projection operators for the MSGs.
An implementation is distributed with a permissive free software license in \software{spglib} v2.0.2: \url{https://github.com/spglib/spglib}.

\end{abstract}





\section{\label{sec:introduction}Introduction}

A crystal symmetry search and the standardization of crystal structures play crucial roles in computational materials science.
For example, symmetry operations are required in irreducible representations of electronic states \cite{GAO2021107760}, band paths \cite{HINUMA2017140}, phonon calculations \cite{phonopy,phono3py}, a random structure search \cite{FREDERICKS2021107810}, and crystal structure description \cite{ganose_jain_2019}.
Moreover, the standardization of crystal structures is indispensable for comparing crystal structures in different settings and analyzing magnetic crystal structures in high-throughput first-principles calculations \cite{Horton2019}.

Owing to the development of a computer-friendly description of space groups \cite{Hall:a19707,shmueli2010symmetry} and algorithms \cite{Opgenorth:js0065,Grosse-Kunstleve:au0146,Grosse-Kunstleve:au0265,Eick2006}, we can automatically perform the crystal symmetry search nowadays.
For example, \textsc{spglib} implements the symmetry-search algorithm and an iterative method to robustly determine crystal symmetries \cite{spglibv1}, which one of the authors has developed and maintained.

On the other hand, algorithms and implementations for magnetic space groups (MSGs) \cite{litvin2016magnetic} remains limited.
MSGs are essential when we consider time-reversal operations or magnetic crystal structures.
To the best of our knowledge, existing implementations only partly provide MSG functionalities.
\software{aflow-sym} \cite{Hicks:ae5042} proposed and implemented a robust space-group analysis algorithm; however, it does not seem to support MSGs yet.
\software{identify magnetic group} \cite{doi:10.1146/annurev-matsci-070214-021008} in the Bilbao Crystallographic Server \cite{aroyo2011crystallography} and \software{CrysFML2008} \cite{RODRIGUEZCARVAJAL199355,Gonzalez-Platas:tu5004} can identify MSGs from magnetic symmetry operations; however, the determination of magnetic symmetry operations from magnetic crystal structures is not supported.
\software{findsym} \cite{https://doi.org/10.1107/S0021889804031528,FINDSYM} supports the determination of magnetic symmetry operations and the identification of MSGs; however, the source code is not freely available.


Here, we present algorithms for determining magnetic symmetry operations of given magnetic crystal structures, identifying magnetic space-group types of given MSGs, searching for transformations to a BNS setting, and symmetrizing the magnetic crystal structures on the basis of the determined MSGs.
Note that the implementation of these algorithms is virtually unattainable without recent development in crystallography:
\citeasnoun{litvin2014magnetic} provided extensive tables for the 1651 MSGs.
\software{iso-mag} \cite{ISOMAG} provides tables of MSGs in both human and computer readable formats.
Magnetic Hall symbols \cite{Gonzalez-Platas:tu5004} and unified (UNI) MSG symbols \cite{Campbell:ib5106} have been developed to represent MSGs or magnetic space-group types unambiguously, which are based on BNS symbols \cite{belov1957neronova,Bradley2009-ze}.
In this paper, we use the magnetic Hall symbols and the MSG datasets tabulated in \citeasnoun{Gonzalez-Platas:tu5004}.
The implementation is distributed under the BSD 3-clause license in \software{spglib} v2.0.2.

This paper is organized as follows.
In Sec.~\ref{sec:msg}, we recall the mathematical structures of MSGs and present definitions and terminology for describing MSGs.
In Sec.~\ref{sec:symmetry_search}, we provide an algorithm for determining magnetic symmetry operations of a given magnetic crystal structure on the basis of equivalence relationships between sites in the magnetic crystal structure.
In Sec.~\ref{sec:identify}, we provide an algorithm to identify a magnetic space-group type of the determined MSG and to search for a transformation from the determined MSG to one in the BNS setting.
In Sec.~\ref{sec:symmetrize}, we provide an algorithm to symmetrize point coordinates and magnetic moments of the magnetic crystal structure from the determined MSG.




\section{\label{sec:msg}Definitions}

Before we discuss algorithms for MSGs and magnetic crystal structures, we describe definitions and terminology for MSGs.
In Sec.~\ref{sec:msg_msg-type}, we define MSGs and derived space groups, which are essential in identifying a magnetic space-group type and searching for a transformation between MSGs.
In Sec.~\ref{sec:msg_type}, we define equivalence relationships between MSGs.
In Sec.~\ref{sec:msg_symbol}, we mention BNS symbols and their settings, which specify representatives of MSGs, and we use them to standardize given MSGs.
Finally, in Sec.~\ref{sec:msg_action}, we give examples of actions of magnetic symmetry operations for magnetic moments.

\subsection{\label{sec:msg_msg-type}MSG and its construct type}



We consider a \term{time-reversal operation} $1'$ and call an index-two group generated from $1'$ as a \term{time-reversal group} $\{ 1, 1' \} \, (\cong \mathbb{Z}_{2})$, where $1$ represents an identity operation.
Let $\mathcal{M}$ be a subgroup of a direct product of three-dimensional Euclidean group $\mathrm{E}(3)$ and $\{ 1, 1' \}$.
An element $(\bm{W}, \bm{w})\theta$ of $\mathcal{M}$ is called a \term{magnetic symmetry operation}, where we call $\bm{W}$ as a matrix part, $\bm{w}$ as a translation part, and $\theta \in \{ 1, 1' \}$ as a time-reversal part of the magnetic symmetry operation.
In particular, $(\bm{W}, \bm{w})1'$ is called an \term{antisymmetry} operation.
A translation subgroup of $\mathcal{M}$ is defined as
\begin{align}
  \mathcal{T}(\mathcal{M}) = \set{ (\bm{E}, \bm{t}) }{ (\bm{E}, \bm{t})1 \in \mathcal{M} },
\end{align}
where $\bm{E}$ represents the identity matrix.
The subgroup $\mathcal{M}$ is called a \term{magnetic space group} (MSG) when $\mathcal{T}(\mathcal{M})$ is generated from three independent translations.
We write a \term{magnetic point group} of $\mathcal{M}$ as
\begin{align}
    \mathcal{P}(\mathcal{M})
        &= \set{ \bm{W}\theta }{ \exists \bm{w} \in \mathbb{R}^{3}, (\bm{W}, \bm{w})\theta \in \mathcal{M} }.
\end{align}

We consider two derived space groups from $\mathcal{M}$.
A \term{family space group} (FSG) of $\mathcal{M}$ is a space group obtained by ignoring time-reversal parts in magnetic symmetry operations:
\begin{align}
    \mathcal{F}(\mathcal{M})
      = \set{ (\bm{W}, \bm{w}) }{ \exists \theta \in \{ 1, 1' \}, (\bm{W}, \bm{w})\theta \in \mathcal{M} }.
\end{align}
A \term{maximal space subgroup} (XSG) of $\mathcal{M}$ is a space group obtained by removing antisymmetry operations:
\begin{align}
    \mathcal{D}(\mathcal{M})
      = \set{ (\bm{W}, \bm{w}) }{ (\bm{W}, \bm{w})1 \in \mathcal{M} }.
\end{align}

The MSGs are classified into the following four \term{construct types} \cite{Bradley2009-ze,Campbell:ib5106}:
\begin{itemize}
  \item (Type I)
    $\mathcal{M} = \mathcal{F}(\mathcal{M})1 = \mathcal{D}(\mathcal{M})1$:
    The MSG $\mathcal{M}$ does not have antisymmetry operations.
  \item (Type II)
    $\mathcal{M} = \mathcal{F}(\mathcal{M})1 \,\sqcup\, \mathcal{F}(\mathcal{M})1', \mathcal{F}(\mathcal{M}) = \mathcal{D}(\mathcal{M})$:
    The MSG $\mathcal{M}$ has antisymmetry operations and corresponding ordinary symmetry operations.
  \item (Type III)
    $\mathcal{M} = \mathcal{D}(\mathcal{M})1 \sqcup (\mathcal{F}(\mathcal{M}) \backslash \mathcal{D}(\mathcal{M})) 1'$ and $\mathcal{D}(\mathcal{M})$ is an index-two \term{translationengleiche} subgroup of $\mathcal{F}(\mathcal{M})$\footnote{
      The notation $\mathcal{F}(\mathcal{M}) \backslash \mathcal{D}(\mathcal{M})$ indicates a complement set, $\mathcal{F}(\mathcal{M}) \backslash \mathcal{D}(\mathcal{M}) = \set{ g\theta \in \mathcal{F}(\mathcal{M}) }{ g\theta \notin \mathcal{D}(\mathcal{M})}$.
    }.
    Thus, translation subgroups of $\mathcal{F}(\mathcal{M})$ and $\mathcal{D}(\mathcal{M})$ are identical.
  \item (Type IV)
    $\mathcal{M} = \mathcal{D}(\mathcal{M})1 \sqcup (\mathcal{F}(\mathcal{M}) \backslash \mathcal{D}(\mathcal{M})) 1'$ and $\mathcal{D}(\mathcal{M})$ is an index-two \term{klassengleiche} subgroup of $\mathcal{F}(\mathcal{M})$.
    Thus, point groups of $\mathcal{F}(\mathcal{M})$ and $\mathcal{D}(\mathcal{M})$ are identical.
\end{itemize}

\begin{figure}
  \centering
  \includegraphics[width=0.95\columnwidth]{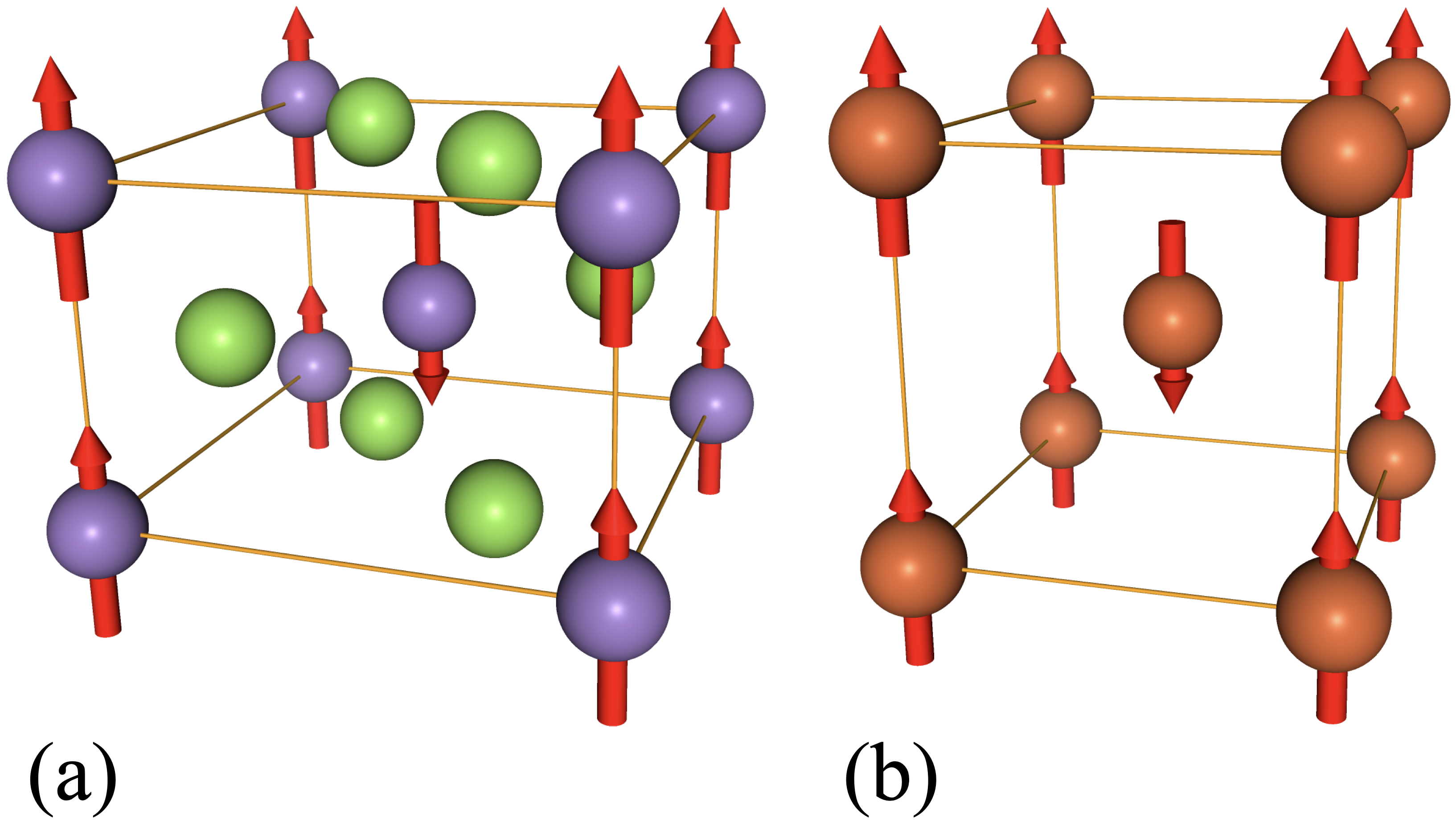}
  \caption{
    Examples of antiferromagnetic (AFM) crystal structures with (a) type-III and (b) type-IV MSGs.
    The red arrows represent collinear spins with the same magnitudes.
  }
  \label{fig:mag_structure_examples}
\end{figure}

For a type-III MSG example, Fig.~\ref{fig:mag_structure_examples}(a) shows an antiferromagnetic (AFM) rutile structure whose MSG is $\mathcal{M}_{\mathrm{rutile}} = P4_{2}'/mn'm$ (BNS number 136.498) in the BNS symbol.
The FSG and XSG of $\mathcal{M}_{\mathrm{rutile}}$ are $P4_{2}/mnm$ (No. 136) and $Pnnm$ (No. 58), respectively.

For a type-IV MSG example, Fig.~\ref{fig:mag_structure_examples}(b) shows an AFM bcc structure whose MSG is $\mathcal{M}_{\mathrm{bcc}} = P_{I}m\overline{3}m$ (BNS number 221.97) in the BNS symbol.
The FSG and XSG of $\mathcal{M}_{\mathrm{bcc}}$ are $Im\overline{3}m$ (No. 229) and $Pm\overline{3}m$ (No. 221), respectively.

\subsection{\label{sec:msg_type}Magnetic space-group type}

We consider a transformation $(\bm{P}, \bm{p})$ between two coordinate systems specified with basis vectors $\bm{A} = (\bm{a}_{1}, \bm{a}_{2}, \bm{a}_{3})$ with origin $\bm{O}$ and basis vectors $\bm{AP}$ with origin $\bm{O} + \bm{Ap}$.
A transformation $(\bm{P}, \bm{p})$ with $\det \bm{P} > 0$ is called orientation-preserving.
We assume that a magnetic symmetry operation $(\bm{W}, \bm{w})\theta$ is transformed into $(\bm{W}', \bm{w}')\theta'$ by $(\bm{P}, \bm{p})$ as
\begin{align}
  (\bm{W}', \bm{w}')
    &= (\bm{P}, \bm{p})^{-1} (\bm{W}, \bm{w}) (\bm{P}, \bm{p}) \\
  \theta' &= \theta.
\end{align}

We refer to the criteria to choose a representative of each space-group type as a \term{setting}.
The \term{standard ITA setting} is one of the conventional descriptions for each space-group type used in the \textit{International Tables for Crystallography Vol. A} \cite{ITA2016}: unique axis b setting, cell choice 1 for monoclinic groups, hexagonal axes for rhombohedral groups, and origin choice 2 for centrosymmetric groups.
Similarly to space groups, each equivalent class of MSGs up to orientation-preserving transformations is called a \term{magnetic space-group type}.


\subsection{\label{sec:msg_symbol}BNS setting}

The BNS symbol represents each magnetic space-group type \cite{belov1957neronova}.
We refer to a setting of the BNS symbol as a \term{BNS setting}: For types-I, -II, and -III MSGs, it uses the same setting as the standard ITA setting of the FSG.
For type-IV MSG, it uses that of the XSG.
In Sec.~\ref{sec:identify}, we consider standardizing a given magnetic crystal structure by applying a transformation to an MSG in the BNS setting.

\subsection{\label{sec:msg_action}Action of magnetic symmetry operations}

In general, we can arbitrarily choose how magnetic symmetry operations act on objects as long as they satisfy the definition of actions.
For a magnetic moment $\bm{m}$, a symmetry operation $(\bm{W}, \bm{w})$ acts on $\bm{m}$ as an axial vector, and the time-reversal operation $1'$ reverses the sign of $\bm{m}$.
When we choose the Cartesian coordinates for $\bm{m}$, the matrix part of $(\bm{W}, \bm{w})$ is expressed as $\bm{A}\bm{W}\bm{A}^{-1}$ in Cartesian coordinates with basis vectors $\bm{A} = (\bm{a}_{1}, \bm{a}_{2}, \bm{a}_{3})$.
Therefore, the magnetic symmetry operations act on $\bm{m}$ as
\begin{align}
  (\bm{W}, \bm{w}) \theta \bm{m}
    &=
    \begin{cases}
      (\mathrm{det} \bm{W}) \bm{A}\bm{W}\bm{A}^{-1} \bm{m} & (\theta = 1) \\
      -(\mathrm{det} \bm{W}) \bm{A}\bm{W}\bm{A}^{-1} \bm{m} & (\theta = 1') \\
    \end{cases}.
\end{align}





\section{\label{sec:symmetry_search}Magnetic symmetry operation search}

We provide a procedure to search for magnetic symmetry operations from a given magnetic crystal structure represented by basis vectors, point coordinates, atomic types, and magnetic moments within a unit cell.
Formally, our input for the magnetic symmetry operation search is the following four objects:
(1) basis vectors of its lattice $\bm{A} = (\bm{a}_{1}, \bm{a}_{2}, \bm{a}_{3})$,
(2) an array of point coordinates of sites in its unit cell $\bm{X} = (\bm{x}_{1}, \cdots, \bm{x}_{N})$,
(3) an array of atomic types of sites in its unit cell $\bm{T} = (t_{1}, \cdots, t_{N})$,
and (4) an array of magnetic moments of sites in its unit cell $\bm{M} = (\bm{m}_{1}, \cdots, \bm{m}_{N})$,
where $N$ is the number of sites in the unit cell.

We search for a magnetic symmetry operation $g\theta \in \mathrm{E}(3) \times \{ 1, 1' \}$ that preserves the magnetic crystal structure $(\bm{A}, \bm{X}, \bm{T}, \bm{M})$.
Therefore, the symmetry operation $g$ should map point coordinates $\bm{x}_{i}$ into $\bm{x}_{\sigma_{g}(i)}$ up to translations, where $\sigma_{g}$ is a permutation of $N$ sites induced by $g$.
Also, $g \theta$ should equate a magnetic moment $\bm{m}_{i}$ with a mapped one $\bm{m}_{\sigma_{g}(i)}$.
Such a magnetic symmetry operation forms an MSG of the magnetic crystal structure as a stabilizer of $\mathrm{E}(3) \times \{ 1, 1' \}$,
\begin{align}
    \label{eq:msg-initial-definition}
    &\mathcal{M}(\bm{A}, \bm{X}, \bm{T}, \bm{M}) \nonumber \\
        &= \mathrm{Stab}_{\mathrm{E}(3) \times \{ 1, 1' \} } \, (\bm{A}, \bm{X}, \bm{T}, \bm{M}) \nonumber \\
        &= \set{
                g\theta \in \mathrm{E}(3) \times \{1, 1' \}
            }{
                \begin{array}{l}
                    \exists \sigma_{g}\in \mathfrak{S}_{N}, \forall i, \\
                    g \bm{x}_{i} \equiv \bm{x}_{\sigma_{g}(i)} \, (\mathrm{mod} \, 1) \\
                    t_{i} = t_{\sigma_{g}(i)}, \\
                    g\theta \bm{m}_{i} = \bm{m}_{\sigma_{g}(i)}
                \end{array}
            },
\end{align}
where $\mathfrak{S}_{N}$ is a symmetric group of degree $N$\footnote{
    We recall that the condition of $\mathcal{M}(\bm{A}, \bm{X}, \bm{T}, \bm{M})$ in Eq.~\eqref{eq:msg-initial-definition} can be read as there exists a permutation $\sigma_{g}$ such that point coordinates, atomic types, and magnetic moments are preserved for every site $i$.
}.

\begin{figure}
  \centering
  \includegraphics[width=0.95\textwidth]{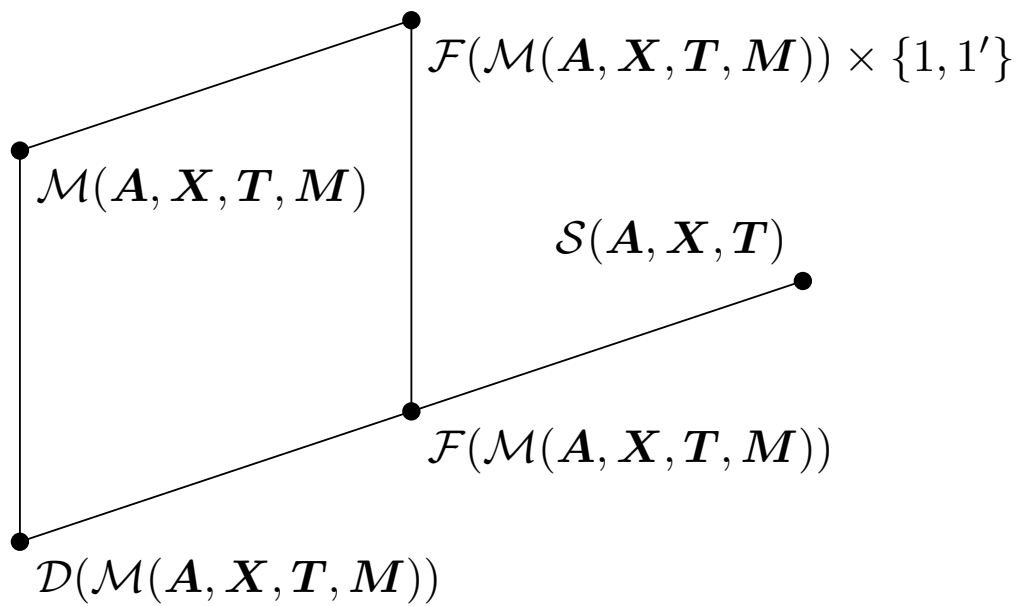}
  \caption{
    Group--subgroup relationship of MSGs and related space groups.
    The nodes represent space groups or MSGs.
    Each edge indicates that a lower group is a subgroup of an upper group in the diagram.
    Although it is not exploited in this study, the XSG $\mathcal{D}(\mathcal{M}(\bm{A}, \bm{X}, \bm{T}, \bm{M}))$ is a subgroup of the FSG $\mathcal{F}(\mathcal{M}(\bm{A}, \bm{X}, \bm{T}, \bm{M}))$ because the latter simply ignores time-reversal parts of $\mathcal{M}(\bm{A}, \bm{X}, \bm{T}, \bm{M})$.
  }
  \label{fig:group_subgroup}
\end{figure}

Because the domain of the symmetry operation $g$ in Eq.~\eqref{eq:msg-initial-definition} is not restricted, we cannot search thoroughly for $g$ at this point.
To narrow down the candidates for $g$, we consider a crystal structure $(\bm{A}, \bm{X}, \bm{T})$ obtained by ignoring the magnetic moments of $(\bm{A}, \bm{X}, \bm{T}, \bm{M})$.
A space group of $(\bm{A}, \bm{X}, \bm{T})$ is written as a stabilizer of $\mathrm{E}(3)$ that preserves $(\bm{A}, \bm{X}, \bm{T})$:
\begin{align}
    \mathcal{S}(\bm{A}, \bm{X}, \bm{T})
        &= \mathrm{Stab}_{\mathrm{E}(3)} \, (\bm{A}, \bm{X}, \bm{T}) \nonumber \\
        &= \set{
                g \in \mathrm{E}(3)
            }{
                \begin{array}{l}
                    \exists \sigma_{g} \in \mathfrak{S}_{N}, \forall i, \\
                    g \bm{x}_{i} \equiv \bm{x}_{\sigma_{g}(i)} \, (\mathrm{mod} \, 1) \\
                    t_{i} = t_{\sigma_{g}(i)}
                \end{array}
            }.
\end{align}
As shown in Fig.~\ref{fig:group_subgroup}, $\mathcal{S}(\bm{A}, \bm{X}, \bm{T})$ may not be a subgroup of $\mathcal{M}(\bm{A}, \bm{X}, \bm{T}, \bm{M})$ in general because the former ignores magnetic moments.
Because time-reversal operations do not change point coordinates and atomic types, we can restrict the domain of symmetry operations $g$ to $\mathcal{S}(\bm{A}, \bm{X}, \bm{T})$,
\begin{align}
    &\label{eq:msg_definition}
    \mathcal{M}(\bm{A}, \bm{X}, \bm{T}, \bm{M}) \nonumber \\
        &= \set{
                g\theta \in \mathcal{S}(\bm{A}, \bm{X}, \bm{T}) \times \{1, 1' \}
            }{
                \begin{array}{l}
                    \exists \sigma_{g}\in \mathfrak{S}_{N}, \forall i, \\
                    g \bm{x}_{i} \equiv \bm{x}_{\sigma_{g}(i)} \, (\mathrm{mod} \, 1) \\
                    t_{i} = t_{\sigma_{g}(i)}, \\
                    g\theta \bm{m}_{i} = \bm{m}_{\sigma_{g}(i)}
                \end{array}
            }.
\end{align}
The symmetry operations for $(\bm{A}, \bm{X}, \bm{T})$ can be obtained from existing crystal symmetry search algorithms such as \citeasnoun{https://doi.org/10.1107/S0021889804031528}, \citeasnoun{spglibv1}, and \citeasnoun{Hicks:ae5042}.

Based on the formulation of the MSG in Eq.~\eqref{eq:msg_definition}, we can search for magnetic symmetry operations using the following procedure:
\begin{enumerate}
    \item We compute $\mathcal{S}(\bm{A}, \bm{X}, \bm{T})$ by the existing crystal symmetry search algorithms.
    \item
        If all magnetic moments are zero, both $g1$ and $g1'$ belong to $\mathcal{M}(\bm{A}, \bm{X}, \bm{T}, \bm{M})$ for all $g \in \mathcal{S}(\bm{A}, \bm{X}, \bm{T})$ and we skip the remaining steps (in this case, the MSG is type II).
        Otherwise, we choose a site $i^{\ast}$ with a non-zero magnetic moment $\bm{m}_{i^{\ast}} (\neq \bm{0})$.
    \item For each symmetry operation $g \in \mathcal{S}(\bm{A}, \bm{X}, \bm{T})$, we search for the time-reversal part as follows:
        \begin{enumerate}
            \item We compute a permutation $\sigma_{g}$ and solve
                \begin{align}
                    \label{eq:time-reversal-part-determination}
                    g \theta \bm{m}_{i^{\ast}} = \bm{m}_{\sigma_{g}(i^{\ast})}
                \end{align}
                for $\theta \in \{ 1, 1' \}$.
                We denote the solution of Eq.~\eqref{eq:time-reversal-part-determination} as $\theta^{\ast}$ if it exists.
                If the solution does not exist, we skip the symmetry operation $g$.
            \item We check if the condition $g \theta^{\ast} \bm{m}_{i} = \bm{m}_{\sigma_{g}(i)}$ holds for other sites.
                If the condition holds for all sites, $g \theta^{\ast}$ belongs to $\mathcal{M}(\bm{A}, \bm{X}, \bm{T}, \bm{M})$.
        \end{enumerate}
\end{enumerate}

Note that the comparison of point coordinates and magnetic moments should be performed within tolerances in practice \cite{Grosse-Kunstleve:sh5006}.
We use an absolute tolerance parameter $\epsilon$ for point coordinates \cite{spglibv1} and another absolute tolerance $\epsilon_{\mathrm{mag}}$ for magnetic moments.
Then, the comparisons in this section are replaced with the following inequalities:
\begin{align}
    g \bm{x}_{i} \equiv \bm{x}_{\sigma_{g}(i)} \, (\mathrm{mod} \, 1)
        &\rightarrow
        \left\lVert \bm{A} \left[ g \bm{x}_{i} - \bm{x}_{\sigma_{g}(i)} \right]_{\mathrm{mod} \, 1} \right\rVert_{2} < \epsilon \\
    g\theta \bm{m}_{i} = \bm{m}_{\sigma_{g}(i)}
        &\rightarrow
        \left\lVert g\theta \bm{m}_{i} - \bm{m}_{\sigma_{g}(i)} \right\rVert_{2} < \epsilon_{\mathrm{mag}}.
\end{align}
Here, $\left[ \cdot \right]_{\mathrm{mod}\, 1}$ takes a remainder with modulo one between $[-0.5, 0.5]$.





\section{\label{sec:identify}Identification of magnetic space-group type and transformation to BNS setting}

For the detected MSG $\mathcal{M} = \mathcal{M}(\bm{A}, \bm{X}, \bm{T}, \bm{M})$ in the previous section, we provide an algorithm to identify its magnetic space-group type and search for a transformation from $\mathcal{M}$ to a magnetic space-group representative $\mathcal{M}_{\mathrm{BNS}}$ in the BNS setting.
The algorithms presented in this section are applied to a list of magnetic symmetry operations in the matrix form either obtained through the magnetic symmetry operation search in Sec.~\ref{sec:symmetry_search} or provided from outside of the software package as predetermined operations.

For all the 1651 magnetic space-group types, a magnetic space-group representative $\mathcal{M}_{\mathrm{BNS}}$ in the BNS setting has already been tabulated \cite{Gonzalez-Platas:tu5004,Campbell:ib5106}.
Thus, we search for $\mathcal{M}_{\mathrm{BNS}}$ with the same magnetic space-group type as $\mathcal{M}$ and an orientation-preserving transformation $(\bm{P}, \bm{p})$ while satisfying
\begin{align}
    \label{eq:matching_msg}
    (\bm{P}, \bm{p})^{-1} \mathcal{M} (\bm{P}, \bm{p}) = \mathcal{M}_{\mathrm{BNS}}.
\end{align}
In Sec.~\ref{sec:identify_construct-type}, we identify a construct type of $\mathcal{M}$ to choose a candidate $\mathcal{M}_{\mathrm{BNS}}$, which is one of the magnetic space-group representatives in the BNS setting.
In Sec.~\ref{sec:identify_comparison}, we try to obtain $(\bm{P}, \bm{p})$ from affine normalizers of $\mathcal{F}(\mathcal{M}_{\mathrm{BNS}})$ or $\mathcal{D}(\mathcal{M}_{\mathrm{BNS}})$.

\subsection{\label{sec:identify_construct-type}Identification of construct type of MSG}

The construct type of $\mathcal{M}$ can be determined from orders of the magnetic point group and point groups of FSG and XSG.
We write a point group of space group $\mathcal{S}$ as
\begin{align}
    \mathcal{P}(\mathcal{S})
        &= \set{ \bm{W} }{ \exists \bm{w} \in \mathbb{R}^{3}, (\bm{W}, \bm{w}) \in \mathcal{S} }.
\end{align}
When $|\mathcal{P}(\mathcal{F}(\mathcal{M}))| / |\mathcal{P}(\mathcal{D}(\mathcal{M}))| = 1 $, $\mathcal{M}$ is type I or II.
Then, when $|\mathcal{P}(\mathcal{M})| / |\mathcal{P}(\mathcal{F}(\mathcal{M}))| = 1$, $\mathcal{M}$ is type I.
When $|\mathcal{P}(\mathcal{M})| / |\mathcal{P}(\mathcal{F}(\mathcal{M}))| = 2$, $\mathcal{M}$ is type II.

When $|\mathcal{P}(\mathcal{F}(\mathcal{M}))| / |\mathcal{P}(\mathcal{D}(\mathcal{M}))| = 2$, $\mathcal{M}$ is type III or IV.
For type-III or type-IV MSG, we consider a coset decomposition of $\mathcal{M}$ by $\mathcal{D}(\mathcal{M})$:
\begin{align}
    \mathcal{M} = \mathcal{D}(\mathcal{M}) 1 \sqcup \mathcal{D}(\mathcal{M}) (\bm{W}_{0}, \bm{w}_{0})1'.
\end{align}
If the coset representative $(\bm{W}_{0}, \bm{w}_{0})1'$ can be taken as an anti-translation, $\mathcal{D}(\mathcal{M})$ is a klassengleiche subgroup of $\mathcal{F}(\mathcal{M})$ and $\mathcal{M}$ is type IV.
If not, $\mathcal{D}(\mathcal{M})$ is a translationengleiche subgroup of $\mathcal{F}(\mathcal{M})$ and $\mathcal{M}$ is type III.

\subsection{\label{sec:identify_comparison}Transformation of MSG to BNS setting}

For each magnetic space-group representative $\mathcal{M}_{\mathrm{BNS}}$ with the same construct type as $\mathcal{M}$, we consider searching for $(\bm{P}, \bm{p})$ from two consecutive transformations $(\bm{P}_{\mathrm{temp}}, \bm{p}_{\mathrm{temp}})$ and $(\bm{P}_{\mathrm{corr}}, \bm{p}_{\mathrm{corr}})$ with
\begin{align}
    \label{eq:two-step-transformation}
    (\bm{P}, \bm{p}) = (\bm{P}_{\mathrm{temp}}, \bm{p}_{\mathrm{temp}}) (\bm{P}_{\mathrm{corr}}, \bm{p}_{\mathrm{corr}})
\end{align}
as described below.
If such a transformation $(\bm{P}, \bm{p})$ is found, $\mathcal{M}$ belongs to the same magnetic space-group type with $\mathcal{M}_{\mathrm{BNS}}$.

We rewrite Eq.~\eqref{eq:matching_msg} to an equivalent one in terms of derived space groups because we would like to use an existing transformation search algorithm to obtain a transformation between space groups with the same space-group type proposed by \citeasnoun{Grosse-Kunstleve:au0146}.
As shown in Appendix~\ref{appx:proof}, the condition of Eq.~\eqref{eq:matching_msg} is equivalent to satisfying the following two conditions:
\begin{subequations}
    \begin{align}
        \label{eq:msg-match-fsg}
        (\bm{P}, \bm{p})^{-1} \mathcal{F}(\mathcal{M}) (\bm{P}, \bm{p}) &= \mathcal{F}(\mathcal{M}_{\mathrm{BNS}}) \\
        \label{eq:msg-match-xsg}
        (\bm{P}, \bm{p})^{-1} \mathcal{D}(\mathcal{M}) (\bm{P}, \bm{p}) &= \mathcal{D}(\mathcal{M}_{\mathrm{BNS}}).
    \end{align}
\end{subequations}
Note that a transformation satisfying Eq.~\eqref{eq:msg-match-fsg} does not necessarily hold Eq.~\eqref{eq:msg-match-xsg} in general, and vice versa.

The present algorithm, based on the new conditions, is outlined as follows.
First, we obtain a temporal transformation $(\bm{P}_{\mathrm{temp}}, \bm{p}_{\mathrm{temp}})$ to match FSGs or XSGs of $\mathcal{M}$ and $\mathcal{M}_{\mathrm{BNS}}$ by the existing transformation search algorithm.
Then, we search for a correction transformation $(\bm{P}_{\mathrm{corr}}, \bm{p}_{\mathrm{corr}})$ to match FSGs and XSGs simultaneously.
We divide the transformation search into cases by the construct type of $\mathcal{M}$ in more detail.

\subsubsection{\label{sec:type1_2_conjugators}When \texorpdfstring{$\mathcal{M}$}{M} is type I or II}

When $\mathcal{M}$ is type I or II, $\mathcal{M}_{\mathrm{BNS}}$ with the same construct type with $\mathcal{M}$ uses the standard ITA setting of $\mathcal{F}(\mathcal{M}_{\mathrm{BNS}})$.
Thus, we need to obtain an orientation-preserving transformation $(\bm{P}_{\mathrm{temp}}, \bm{p}_{\mathrm{temp}})$ such that $(\bm{P}_{\mathrm{temp}}, \bm{p}_{\mathrm{temp}})^{-1} \mathcal{F}(\mathcal{M}) (\bm{P}_{\mathrm{temp}}, \bm{p}_{\mathrm{temp}}) = \mathcal{F}(\mathcal{M}_{\mathrm{BNS}})$.
The temporal transformation $(\bm{P}_{\mathrm{temp}}, \bm{p}_{\mathrm{temp}})$ can be obtained by the existing transformation search algorithm.
We write an MSG transformed by $(\bm{P}_{\mathrm{temp}}, \bm{p}_{\mathrm{temp}})$ as
\begin{align}
    \label{eq:msg_half_matched}
    \mathcal{M}_{\mathrm{temp}} = (\bm{P}_{\mathrm{temp}}, \bm{p}_{\mathrm{temp}})^{-1} \mathcal{M} (\bm{P}_{\mathrm{temp}}, \bm{p}_{\mathrm{temp}}).
\end{align}
By construction, $\mathcal{F}(\mathcal{M}_{\mathrm{temp}})$ and $\mathcal{F}(\mathcal{M}_{\mathrm{BNS}})$ are identical as sets, $\mathcal{F}(\mathcal{M}_{\mathrm{temp}}) = \mathcal{F}(\mathcal{M}_{\mathrm{BNS}})$.

In this case, the XSGs are also identical to one another, $\mathcal{D}(\mathcal{M}_{\mathrm{temp}}) = \mathcal{F}(\mathcal{M}_{\mathrm{temp}}) = \mathcal{F}(\mathcal{M}_{\mathrm{BNS}}) = \mathcal{D}(\mathcal{M}_{\mathrm{BNS}})$.
Thus, we do not need to search for a correction transformation because $(\bm{P}_{\mathrm{temp}}, \bm{p}_{\mathrm{temp}})$ also satisfies $(\bm{P}_{\mathrm{temp}}, \bm{p}_{\mathrm{temp}})^{-1} \mathcal{D}(\mathcal{M}) (\bm{P}_{\mathrm{temp}}, \bm{p}_{\mathrm{temp}}) = \mathcal{D}(\mathcal{M}_{\mathrm{BNS}})$.

\subsubsection{\label{sec:type3_conjugators}When \texorpdfstring{$\mathcal{M}$}{M} is type III}

When $\mathcal{M}$ is type III, $\mathcal{M}_{\mathrm{BNS}}$ with type III uses the standard ITA setting of $\mathcal{F}(\mathcal{M}_{\mathrm{BNS}})$.
Thus, we need to obtain an orientation-preserving transformation $(\bm{P}_{\mathrm{temp}}, \bm{p}_{\mathrm{temp}})$ such that $(\bm{P}_{\mathrm{temp}}, \bm{p}_{\mathrm{temp}})^{-1} \mathcal{F}(\mathcal{M}) (\bm{P}_{\mathrm{temp}}, \bm{p}_{\mathrm{temp}}) = \mathcal{F}(\mathcal{M}_{\mathrm{BNS}})$.
Then, the FSG of the transformed MSG in Eq.~\eqref{eq:msg_half_matched}, $\mathcal{F}(\mathcal{M}_{\mathrm{temp}})$, is the space-group representative in the standard ITA setting.

\begin{figure}
  \centering
  \includegraphics[width=0.95\textwidth]{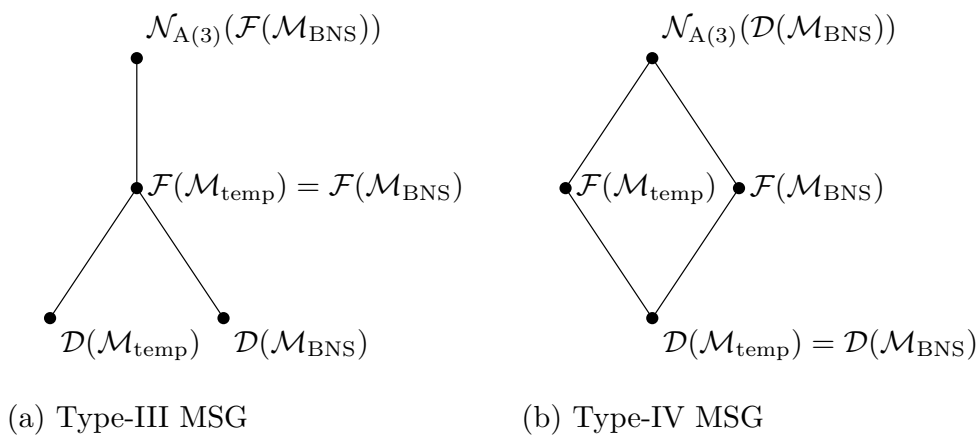}
  \caption{Group--subgroup relationship of conjugated MSGs and affine normalizers.}
  \label{fig:alternatives}
\end{figure}

A correction transformation $(\bm{P}_{\mathrm{corr}}, \bm{p}_{\mathrm{corr}})$ should satisfy the following conditions to simultaneously hold Eqs.~\eqref{eq:msg-match-fsg} and \eqref{eq:msg-match-xsg},
\begin{subequations}
    \begin{align}
        \label{eq:conjugated_fsg_normalizer}
        (\bm{P}_{\mathrm{corr}}, \bm{p}_{\mathrm{corr}})^{-1} \mathcal{F}(\mathcal{M}_{\mathrm{BNS}}) (\bm{P}_{\mathrm{corr}}, \bm{p}_{\mathrm{corr}}) &= \mathcal{F}(\mathcal{M}_{\mathrm{BNS}})\\
        \label{eq:conjugated_xsg}
        (\bm{P}_{\mathrm{corr}}, \bm{p}_{\mathrm{corr}})^{-1} \mathcal{D}(\mathcal{M}_{\mathrm{temp}}) (\bm{P}_{\mathrm{corr}}, \bm{p}_{\mathrm{corr}}) &= \mathcal{D}(\mathcal{M}_{\mathrm{BNS}}).
    \end{align}
\end{subequations}
The condition of Eq.~\eqref{eq:conjugated_fsg_normalizer} indicates that $(\bm{P}_{\mathrm{corr}}, \bm{p}_{\mathrm{corr}})$ belongs to an affine normalizer of $\mathcal{F}(\mathcal{M}_{\mathrm{BNS}})$ \cite{koch2016normalizers}.
The situation is shown in Fig.~\ref{fig:alternatives}(a), where we write the affine normalizer of a space group $\mathcal{S}$ as
\begin{align}
    \mathcal{N}_{\mathrm{A}(3)}(\mathcal{S})
    = \set{
        (\bm{Q}, \bm{q}) \in \mathrm{A}(3)
    } {
        (\bm{Q}, \bm{q})^{-1} \mathcal{S} (\bm{Q}, \bm{q}) = \mathcal{S}
    }
\end{align}
and the three-dimensional affine group as $\mathrm{A}(3)$.
If a correction transformation $(\bm{P}_{\mathrm{corr}}, \bm{p}_{\mathrm{corr}}) \in \mathcal{N}_{\mathrm{A}(3)}(\mathcal{F}(\mathcal{M}_{\mathrm{BNS}}))$ satisfies Eq.~\eqref{eq:conjugated_xsg}, the combined transformation in Eq.~\eqref{eq:two-step-transformation} transforms $\mathcal{M}$ to $\mathcal{M}_{\mathrm{BNS}}$.

Finally, we describe how to prepare transformations in the affine normalizer $\mathcal{N}_{\mathrm{A}(3)}(\mathcal{F}(\mathcal{M}_{\mathrm{BNS}}))$ in practice.
Because $\mathcal{D}(\mathcal{M}_{\mathrm{BNS}})$ is a normal subgroup of $\mathcal{F}(\mathcal{M}_{\mathrm{BNS}})$, an operation in $\mathcal{F}(\mathcal{M}_{\mathrm{BNS}})$ does not give another conjugated subgroup of $\mathcal{D}(\mathcal{M}_{\mathrm{BNS}})$.
Also, although the affine normalizer may have continuous translations, the continuous translations do not give another conjugated subgroup of $\mathcal{D}(\mathcal{M}_{\mathrm{BNS}})$.
Thus, it is sufficient to consider coset representatives of $\mathcal{N}_{\mathrm{A}(3)}(\mathcal{F}(\mathcal{M}_{\mathrm{BNS}})) / \mathcal{F}(\mathcal{M}_{\mathrm{BNS}})$ other than continuous translations.
We divide the affine normalizer computation into cases whether the number of the coset representatives other than continuous translations is finite or not.

When $\mathcal{F}(\mathcal{M}_{\mathrm{BNS}})$ is triclinic or monoclinic, the number of the coset representatives other than continuous translations is infinite and we cannot check transformations thoroughly.
However, there are no such conjugate space groups with $\mathcal{D}(\mathcal{M}_{\mathrm{temp}}) \neq \mathcal{D}(\mathcal{M}_{\mathrm{BNS}})$ because $\mathcal{P}(\mathcal{F}(\mathcal{M}_{\mathrm{BNS}}))$ does not have a pair of proper conjugate subgroups in its affine normalizer.
Therefore, we do not need to compute the affine normalizer in this case.

When $\mathcal{F}(\mathcal{M}_{\mathrm{BNS}})$ belongs to other crystal systems, the number of the coset representatives other than continuous translations is finite.
To simplify the present algorithm and implementation, instead of using a list of affine normalizers in \citeasnoun{koch2016normalizers}, we enumerate matrix parts and origin shifts of orientation-preserving transformations in the coset representatives other than continuous translations as follows.
For matrix parts, we enumerate integer matrices $\bm{P}_{\mathrm{corr}}$ whose elements are -1, 0, or 1, and their determinants are equal to one.
For origin shifts, we enumerate vectors $\bm{p}_{\mathrm{corr}}$ by restricting their vector components to one of $\left\{ 0, \frac{1}{4}, \frac{1}{3}, \frac{1}{2}, \frac{2}{3}, \frac{3}{4} \right\}$.
These will be sufficient because they cover all orientation-preserving coset representatives of $\mathcal{N}_{\mathrm{A}(3)}(\mathcal{F}(\mathcal{M}_{\mathrm{BNS}})) / \mathcal{F}(\mathcal{M}_{\mathrm{BNS}})$ up to translations \cite{koch2016normalizers}.
Since $(\bm{P}_{\mathrm{corr}}, \bm{p}_{\mathrm{corr}})$ can be tabulated for each space-group representative in the standard ITA setting, we can precompute them before performing the transformation search in practice.


\subsubsection{When \texorpdfstring{$\mathcal{M}$}{M} is type IV}

When $\mathcal{M}$ is type IV, $\mathcal{M}_{\mathrm{BNS}}$ with type IV uses the standard ITA setting of $\mathcal{D}(\mathcal{M}_{\mathrm{BNS}})$.
Thus, we need to obtain an orientation-preserving transformation $(\bm{P}_{\mathrm{temp}}, \bm{p}_{\mathrm{temp}})$ such that $(\bm{P}_{\mathrm{temp}}, \bm{p}_{\mathrm{temp}})^{-1} \mathcal{D}(\mathcal{M}) (\bm{P}_{\mathrm{temp}}, \bm{p}_{\mathrm{temp}}) = \mathcal{D}(\mathcal{M}_{\mathrm{BNS}})$.
Then, the XSG of the transformed MSG in Eq.~\eqref{eq:msg_half_matched}, $\mathcal{D}(\mathcal{M}_{\mathrm{temp}})$, is the space-group representative in the standard ITA setting.

Similarly to type-III MSGs, we need to search for an orientation-preserving transformation $(\bm{P}_{\mathrm{corr}}, \bm{p}_{\mathrm{corr}}) \in \mathcal{N}_{\mathrm{A}(3)}(\mathcal{D}(\mathcal{M}_{\mathrm{BNS}}))$ such that
\begin{align}
    \label{eq:conjugated_fsg}
    (\bm{P}_{\mathrm{corr}}, \bm{p}_{\mathrm{corr}})^{-1} \mathcal{F}(\mathcal{M}_{\mathrm{temp}}) (\bm{P}_{\mathrm{corr}}, \bm{p}_{\mathrm{corr}}) = \mathcal{F}(\mathcal{M}_{\mathrm{BNS}}).
\end{align}
The situation is shown in Fig.~\ref{fig:alternatives}(b) \footnote{
    The FSG $\mathcal{F}(\mathcal{M}_{\mathrm{BNS}})$ is a subgroup of $\mathcal{N}_{\mathrm{A}(3)}(\mathcal{D}(\mathcal{M}_{\mathrm{BNS}}))$: because $\mathcal{D}(\mathcal{M}_{\mathrm{BNS}})$ is a normal subgroup of $\mathcal{F}(\mathcal{M}_{\mathrm{BNS}})$, every operation in $\mathcal{F}(\mathcal{M}_{\mathrm{BNS}})$ stabilizes $\mathcal{D}(\mathcal{M}_{\mathrm{BNS}})$ and belongs to $\mathcal{N}_{\mathrm{A}(3)}(\mathcal{D}(\mathcal{M}_{\mathrm{BNS}}))$.
    Similarly, $\mathcal{F}(\mathcal{M}_{\mathrm{temp}})$ is a subgroup of $\mathcal{N}_{\mathrm{A}(3)}(\mathcal{D}(\mathcal{M}_{\mathrm{BNS}}))$.
}.

When $\mathcal{D}(\mathcal{M}_{\mathrm{BNS}})$ is neither triclinic nor monoclinic, the brute-force tabulation in Sec.~\ref{sec:type3_conjugators} also works for $\mathcal{N}_{\mathrm{A}(3)}(\mathcal{D}(\mathcal{M}_{\mathrm{BNS}}))$.
For triclinic and monoclinic cases, a factor group $\mathcal{N}_{\mathrm{A}(3)}(\mathcal{D}(\mathcal{M}_{\mathrm{BNS}})) / \mathcal{D}(\mathcal{M}_{\mathrm{BNS}})$ is not finite, and we cannot prove the completeness in the same manner.
Thus, we show that the enumerated $(\bm{P}_{\mathrm{corr}}, \bm{p}_{\mathrm{corr}})$ covers all conjugated type-IV MSGs by explicitly listing $(\bm{P}_{\mathrm{corr}}, \bm{p}_{\mathrm{corr}})$ and the conjugated MSGs in Appendix~\ref{appx:proof_type4}.

\subsubsection{Examples of conjugated MSGs}

We present examples of conjugated MSGs for type III and type IV.
For a type-III MSG example, consider coset representatives of $\mathcal{M}_{\mathrm{BNS}}$ for $P22'2_{1}'$ in the BNS setting (BNS number 17.10) as follows:
\begin{align*}
    &x,y,z,1;
    x,-y,-z,1; \\
    &-x,-y,z+1/2,1';
    -x,y,-z+1/2,1'.
\end{align*}
There is another MSG $\mathcal{M}_{\mathrm{temp}}$ with the same magnetic space-group type as $\mathcal{M}_{\mathrm{BNS}}$ and identical FSG to $\mathcal{M}_{\mathrm{BNS}}$:
\begin{align*}
    &x,y,z,1;
    -x,y,-z+1/2,1; \\
    &-x,-y,z+1/2,1';
    x,-y,-z,1'.
\end{align*}
Although $\mathcal{D}(\mathcal{M}_{\mathrm{BNS}})$ and $\mathcal{D}(\mathcal{M}_{\mathrm{temp}})$ belong to the same space-group type (No. 3), these XSGs are different.
The following transformation maps $\mathcal{M}_{\mathrm{temp}}$ to $\mathcal{M}_{\mathrm{BNS}}$ while satisfying Eq.~\eqref{eq:conjugated_xsg}:
\begin{align*}
    (\bm{P}_{\mathrm{corr}}, \bm{p}_{\mathrm{corr}})
    = \left(
        \begin{pmatrix}
            0  & -1 & 0  \\
            -1 & 0  & 0  \\
            0  & 0  & -1 \\
        \end{pmatrix},
        \begin{pmatrix}
            0 \\
            0 \\
            \frac{1}{4} \\
        \end{pmatrix}
    \right).
\end{align*}

For a type-IV MSG example, consider coset representatives of $\mathcal{M}_{\mathrm{BNS}}$ for $C_{c}c$ in the BNS setting (BNS number 9.40) as follows:
\begin{align*}
    &x,y,z,1;
    x,-y,z+1/2,1; \\
    &x+1/2,y+1/2,z,1;
    x+1/2,-y+1/2,z+1/2,1; \\
    &x+1/2,-y+1/2,z,1';
    x,-y,z,1'; \\
    &x,y,z+1/2,1';
    x+1/2,y+1/2,z+1/2,1'.
\end{align*}
There is another MSG $\mathcal{M}_{\mathrm{temp}}$ with the same magnetic space-group type as $\mathcal{M}_{\mathrm{BNS}}$ and identical XSG to $\mathcal{M}_{\mathrm{BNS}}$:
\begin{align*}
    &x,y,z,1;
    x+1/2,-y+1/2,z+1/2,1; \\
    &x+1/2,y+1/2,z,1;
    x,-y,z+1/2,1; \\
    &x+1/2,-y,z,1';
    x,-y+1/2,z,1'; \\
    &x+1/2,y,z+1/2,1';
    x,y+1/2,z+1/2,1'.
\end{align*}
Although $\mathcal{F}(\mathcal{M}_{\mathrm{BNS}})$ and $\mathcal{F}(\mathcal{M}_{\mathrm{temp}})$ belong to the same space-group type (No. 8), these FSGs are different.
The following transformation maps $\mathcal{M}_{\mathrm{temp}}$ to $\mathcal{M}_{\mathrm{BNS}}$ while satisfying Eq.~\eqref{eq:conjugated_fsg}:
\begin{align*}
    (\bm{P}_{\mathrm{corr}}, \bm{p}_{\mathrm{corr}})
    = \left(
        \begin{pmatrix}
            -1 & 0  & -1 \\
            0  & -1 & 0  \\
            0  & 0  & 1  \\
        \end{pmatrix},
        \begin{pmatrix}
            0 \\
            \frac{1}{4} \\
            0 \\
        \end{pmatrix}
    \right).
\end{align*}



\section{\label{sec:symmetrize}Symmetrization of magnetic crystal structure}

We symmetrize the magnetic crystal structure $(\bm{A}, \bm{X}, \bm{T}, \bm{M})$ by magnetic symmetry operations of the determined MSG $\mathcal{M}(\bm{A}, \bm{X}, \bm{T}, \bm{M})$.
For convenience, we consider its coset decomposition with a finite index as follows.
Let $\mathcal{T}_{\bm{A}}$ be a translation group formed by basis vectors $\bm{A}$, which may not be primitive basis vectors.
We write a coset decomposition of $\mathcal{M}(\bm{A}, \bm{X}, \bm{T}, \bm{M})$ by $\mathcal{T}_{\bm{A}}$ as
\begin{align}
    \mathcal{M}(\bm{A}, \bm{X}, \bm{T}, \bm{M})
    &= \bigsqcup_{\kappa} (\bm{W}_{\kappa}, \bm{w}_{\kappa}) \theta_{\kappa} \mathcal{T}_{\bm{A}}.
\end{align}
We write the set of coset representatives as
\begin{align}
    \overline{\mathcal{M}} = \{ (\bm{W}_{\kappa}, \bm{w}_{\kappa}) \theta_{\kappa} \}_{\kappa}.
\end{align}
A centering operation $(\bm{E}, \bm{w})1$, where $\bm{w} \not\equiv \bm{0} \, (\mathrm{mod}\, 1)$, may belong to $\overline{\mathcal{M}}$.

A procedure to symmetrize the array of point coordinates $\bm{X}$ by $\overline{\mathcal{M}}$ is essentially the same as those used by \citeasnoun{Grosse-Kunstleve:au0265} and \citeasnoun{spglibv1}.
For the $\kappa$th magnetic symmetry operation $(\bm{W}_{\kappa}, \bm{w}_{\kappa}) \theta_{\kappa}$, we denote that its inverse maps the $\sigma_{\kappa}^{-1}(i)$th point coordinates to the $i$th point coordinates.
Then, $(\bm{W}_{\kappa}, \bm{w}_{\kappa}) \bm{x}_{ \sigma_{\kappa}^{-1}(i) }$ should be close to $\bm{x}_{i}$ up to lattice translations in $\mathcal{T}_{\bm{A}}$.
With this observation, each of the point coordinates $\bm{x}_{i}$ can be symmetrized to $\tilde{\bm{x}}_{i}$ by a projection operator:
\begin{align}
    \tilde{\bm{x}}_{i}
    =
    \bm{x}_{i}
    + \frac{1}{| \overline{\mathcal{M}} |}
        \sum_{\kappa}
        \left[
            (\bm{W}_{\kappa}, \bm{w}_{\kappa}) \bm{x}_{ \sigma_{\kappa}^{-1}(i) } - \bm{x}_{i}
        \right]_{\mathrm{mod}\, 1}.
\end{align}
The modulo is required because the original and mapped point coordinates in the unit cell may be displaced by lattice translations.

A procedure to symmetrize the array of magnetic moments $\bm{M}$ is similar to that to symmetrize the array of point coordinates.
Each magnetic moment $\bm{m}_{i}$ can be symmetrized to $\tilde{\bm{m}}_{i}$ by the following projection operator:
\begin{align}
    \tilde{\bm{m}}_{i}
    =
    \frac{1}{| \overline{\mathcal{M}} |} \sum_{\kappa} (\bm{W}_{\kappa}, \bm{w}_{\kappa})\theta_{\kappa} \bm{m}_{ \sigma_{\kappa}^{-1}(i) }.
\end{align}



\section{\label{sec:conclusion}Conclusion}

We have presented the algorithms for determining magnetic symmetry operations for a given magnetic crystal structure, identifying a magnetic space-group type for a given MSG, searching for a transformation to the BNS setting, and symmetrizing the magnetic crystal structure on the basis of the determined MSG.
Matrix and translation parts of magnetic symmetry operations are determined from the crystal structure by ignoring magnetic moments.
A transformation between the determined MSG and a BNS-setting MSG is obtained by considering affine normalizers: that of the FSG for type-III MSGs and that of the XSG for type-IV MSGs.
In particular, we provide exhaustive tables of conjugated MSGs with triclinic or monoclinic type-IV MSGs in the BNS setting and corresponding transformations.
Projection operators of the determined MSG symmetrize point coordinates and magnetic moments of the magnetic crystal structure.
These algorithms are designed comprehensively and implemented in \software{spglib} under the BSD 3-clause license.
The present algorithms and their implementations are expected to contribute to computational crystallography and materials science, including high-throughput first-principles calculations and crystal structure predictions.




\appendix


\section{\label{appx:proof}Condition that two MSGs are identical}

For two MSGs $\mathcal{M}_{1}$ and $\mathcal{M}_{2}$, we write the FSG and XSG of $\mathcal{M}_{i} \, (i=1, 2)$ as $\mathcal{F}_{i}$ and $\mathcal{D}_{i}$, respectively.
When $\mathcal{M}_{1}$ and $\mathcal{M}_{2}$ have the same construct types, they are identical if and only if their FSGs and XSGs are also identical, that is, $\mathcal{F}_{1} = \mathcal{F}_{2}$ and $\mathcal{D}_{1} = \mathcal{D}_{2}$.
Although it is trivial, we give proof of this fact for completeness.

When $\mathcal{M}_{1}$ and $\mathcal{M}_{2}$ are identical, their FSGs and XSGs are also identical by definition.
We check the converse for each construct type.
For type-I MSGs, $\mathcal{M}_{1} = \mathcal{F}_{1} 1 = \mathcal{F}_{2} 1 = \mathcal{M}_{2}$.
For type-II MSGs, $\mathcal{M}_{1} = \mathcal{F}_{1} 1 \sqcup \mathcal{F}_{1} 1' = \mathcal{F}_{2} 1 \sqcup \mathcal{F}_{2} 1' = \mathcal{M}_{2}$.
For type-III or type-IV MSGs, $\mathcal{M}_{1} = \mathcal{D}_{1} 1 \sqcup \left( \mathcal{F}_{1} \backslash \mathcal{D}_{1} \right) 1' = \mathcal{D}_{2} 1 \sqcup \left( \mathcal{F}_{2} \backslash \mathcal{D}_{2} \right) 1' = \mathcal{M}_{2}$.
Thus, if FSGs and XSGs are identical, the two MSGs are also identical.

\section{\label{appx:proof_type4}Correction transformations for triclinic or monoclinic type-IV MSGs}

We give all anti-translations in conjugated MSGs and corresponding transformations $(\bm{P}, \bm{p})$ for a type-IV MSG $\mathcal{M}_{\mathrm{BNS}}$ in the BNS setting, where $\mathcal{D}(\mathcal{M}_{\mathrm{BNS}})$ is triclinic or monoclinic.
Because an anti-translation $(\bm{E}, \bm{w})1'$ in a type-IV MSG is index-two up to translations, it is sufficient to consider the following seven anti-translations:
\begin{align}
    1'_{a} = \left( \bm{E}, \left( \frac{1}{2}, 0, 0 \right) \right) 1' \\
    1'_{b} = \left( \bm{E}, \left( 0, \frac{1}{2}, 0 \right) \right) 1' \\
    1'_{c} = \left( \bm{E}, \left( 0, 0, \frac{1}{2} \right) \right) 1' \\
    1'_{bc} = \left( \bm{E}, \left( 0, \frac{1}{2}, \frac{1}{2} \right) \right) 1' \\
    1'_{ac} = \left( \bm{E}, \left( \frac{1}{2}, 0, \frac{1}{2} \right) \right) 1' \\
    1'_{ab} = \left( \bm{E}, \left( \frac{1}{2}, \frac{1}{2}, 0 \right) \right) 1' \\
    1'_{abc} = \left( \bm{E}, \left( \frac{1}{2}, \frac{1}{2}, \frac{1}{2} \right) \right) 1'.
\end{align}

When $\mathcal{D}(\mathcal{M}_{\mathrm{BNS}})$ is a triclinic or monoclinic $P$-centering space group (Nos. 1, 2, 3, 4, 6, 7, 10, 11, 13, and 14), Tables~\ref{tab:type4_conjugators_1_2}, \ref{tab:type4_conjugators_3_4_6_10_11}, and \ref{tab:type4_conjugators_7_13_14} show transformations for $\mathcal{M}_{\mathrm{BNS}}$, which are obtained by the brute force described in Sec.~\ref{sec:type3_conjugators}, and anti-translations in the transformed MSGs.
Because each table contains the seven anti-translations, we confirm that these transformations are sufficient to search for conjugated type-IV MSGs.

For other cases, $\mathcal{D}(\mathcal{M}_{\mathrm{BNS}})$ is a monoclinic $C$-centering space group (Nos. 5, 8, 9, 12, and 15).
Tables~\ref{tab:type4_conjugators_5_8_12}, \ref{tab:type4_conjugators_9}, and \ref{tab:type4_conjugators_15} show transformations for $\mathcal{M}_{\mathrm{BNS}}$ and anti-translations in the transformed MSGs.
Note that the anti-translation $1_{ab}'$ should not be contained in the conjugated MSGs because $\mathcal{M}_{\mathrm{BNS}}$ is not type II and $\mathcal{D}(\mathcal{M}_{\mathrm{BNS}})$ is $C$-centering.
Then, each table contains the six anti-translations other than $1_{ab}'$.
Thus, we also confirm that these transformations are sufficient.






\ack{We would like to thank Juan Rodr\'{i}guez-Carvajal for providing magnetic space-group datasets tabulated in \citeasnoun{Gonzalez-Platas:tu5004}.}


\referencelist[references] 


\begin{table}
  \caption{Transformations between type-IV MSG $\mathcal{M}_{\mathrm{BNS}}$ and conjugated MSGs, where their XSGs are identical to space groups with types Nos. 1 and 2 in the ITA standard setting.}
  \label{tab:type4_conjugators_1_2}
  \begin{tabular}{c|cc}
    \hline
    BNS number & Transformation $(\bm{P}, \bm{p})$ & Anti-translations in $(\bm{P}, \bm{p})^{-1}\mathcal{M}_{\mathrm{BNS}}(\bm{P}, \bm{p})$\\ \hline
    1.3, 2.7
      & $(\bm{a}, \bm{b}, \bm{c}; 0, 0, 0)$ & $1_{c}'$ \\
      & $(-\bm{c}, \bm{a}+\bm{b}+\bm{c}, \bm{a}+\bm{c}; 0, 0, 0)$ & $1_{a}'$ \\
      & $(\bm{a}+\bm{b}+\bm{c}; \bm{c}, \bm{a}+\bm{c}; 0, 0, 0)$ & $1_{b}'$ \\
      & $(\bm{a}+\bm{b}+\bm{c}, -\bm{a}, \bm{a}+\bm{c}; 0, 0, 0)$ & $1_{bc}'$ \\
      & $(\bm{a}+\bm{b}, -\bm{a}-\bm{c}, \bm{a}+\bm{b}+\bm{c}; 0, 0, 0)$ & $1_{ac}'$ \\
      & $(\bm{a}+\bm{b}+\bm{c}, \bm{a}+\bm{b}, -\bm{a}-\bm{c}; 0, 0, 0)$ & $1_{ab}'$ \\
      & $(\bm{a}+\bm{b}+\bm{c}, \bm{a}+\bm{c}, -\bm{b}-\bm{c}; 0, 0, 0)$ & $1_{abc}'$ \\
    \hline
  \end{tabular}
\end{table}

\begin{table}
  \caption{Transformations between type-IV MSG $\mathcal{M}_{\mathrm{BNS}}$ and conjugated MSGs, where their XSGs are identical to space groups with types Nos. 3, 4, 6, 10, and 11 in the ITA standard setting.}
  \label{tab:type4_conjugators_3_4_6_10_11}
  \begin{tabular}{c|cc}
    \hline
    BNS number & Transformation $(\bm{P}, \bm{p})$ & Anti-translations in $(\bm{P}, \bm{p})^{-1}\mathcal{M}_{\mathrm{BNS}}(\bm{P}, \bm{p})$\\ \hline
    3.4, 4.10, 6.21, 10.47, 11.55
      & $(\bm{a}, \bm{b}, \bm{c}; 0, 0, 0)$ & $1_{a}'$ \\
      & $(\bm{a}+\bm{c}, \bm{b}, -\bm{a}; 0, 0, 0)$ & $1_{c}'$ \\
      & $(\bm{a}+\bm{c}, \bm{b}, \bm{c}; 0, 0, 0)$ & $1_{ac}'$ \\
    3.5, 4.11, 6.22, 10.48, 11.56
      & $(\bm{a}, \bm{b}, \bm{c}; 0, 0, 0)$ & $1_{b}'$ \\
    3.6, 4.12, 6.23, 10.49, 11.57
      & $(\bm{a}, \bm{b}, \bm{c}; 0, 0, 0)$ & $1_{ab}'$ \\
      & $(\bm{a}+\bm{c}, \bm{b}, -\bm{a}; 0, 0, 0)$ & $1_{bc}'$ \\
      & $(\bm{a}+\bm{c}, \bm{b}, \bm{c}; 0, 0, 0)$ & $1_{abc}'$ \\
    \hline
  \end{tabular}
\end{table}

\begin{table}
  \caption{Transformations between type-IV MSG $\mathcal{M}_{\mathrm{BNS}}$ and conjugated MSGs, where their XSGs are identical to space groups with types Nos. 5, 8, and 12 in the ITA standard setting.}
  \label{tab:type4_conjugators_5_8_12}
  \begin{tabular}{c|cc}
    \hline
    BNS number & Transformation $(\bm{P}, \bm{p})$ & Anti-translations in $(\bm{P}, \bm{p})^{-1}\mathcal{M}_{\mathrm{BNS}}(\bm{P}, \bm{p})$\\ \hline
    5.16, 8.35, 12.63
      & $(\bm{a}, \bm{b}, \bm{c}; 0, 0, 0)$ & $1_{c}'$, $1_{abc}'$ \\
      & $(\bm{a}, \bm{b}, -\bm{a}+\bm{c}; 0, 0, 0)$ & $1_{ac}'$, $1_{bc}'$ \\
    5.17, 8.36, 12.64
      & $(\bm{a}, \bm{b}, \bm{c}; 0, 0, 0)$ & $1_{a}'$, $1_{b}'$ \\
    \hline
  \end{tabular}
\end{table}

\begin{table}
  \caption{
    Transformations between type-IV MSG $\mathcal{M}_{\mathrm{BNS}}$ and conjugated MSGs, where their XSGs are identical to space groups with types Nos. 7, 13, and 14 in the ITA standard setting.
    Note that BNS numbers 7.30 and 7.31 are not listed in ascending order.
  }
  \label{tab:type4_conjugators_7_13_14}
  \begin{tabular}{c|cc}
    \hline
    BNS number & Transformation $(\bm{P}, \bm{p})$ & Anti-translations in $(\bm{P}, \bm{p})^{-1}\mathcal{M}_{\mathrm{BNS}}(\bm{P}, \bm{p})$\\ \hline
    7.27, 13.70, 14.80
      & $(\bm{a}, \bm{b}, \bm{c}; 0, 0, 0)$ & $1_{a}'$ \\
      & $(\bm{a}+\bm{c}, \bm{b}, \bm{c}; 0, 0, 0)$ & $1_{ac}'$ \\
    7.28, 13.71, 14.81
      & $(\bm{a}, \bm{b}, \bm{c}; 0, 0, 0)$ & $1_{b}'$ \\
    7.29, 13.72, 14.82
      & $(\bm{a}, \bm{b}, \bm{c}; 0, 0, 0)$ & $1_{c}'$ \\
    7.31, 13.73, 14.83
      & $(\bm{a}, \bm{b}, \bm{c}; 0, 0, 0)$ & $1_{bc}'$ \\
    7.30, 13.74, 14.84
      & $(\bm{a}, \bm{b}, \bm{c}; 0, 0, 0)$ & $1_{ab}'$ \\
      & $(\bm{a}+\bm{c}, \bm{b}, \bm{c}; 0, 0, 0)$ & $1_{abc}'$ \\
    \hline
  \end{tabular}
\end{table}

\begin{table}
  \caption{Transformations between type-IV MSG $\mathcal{M}_{\mathrm{BNS}}$ and conjugated MSGs, where their XSGs are identical to a space group with type No. 9 in the ITA standard setting.}
  \label{tab:type4_conjugators_9}
  \begin{tabular}{c|cc}
    \hline
    BNS number & Transformation $(\bm{P}, \bm{p})$ & Anti-translations in $(\bm{P}, \bm{p})^{-1}\mathcal{M}_{\mathrm{BNS}}(\bm{P}, \bm{p})$\\ \hline
    9.40
      & $(\bm{a}, \bm{b}, \bm{c}; 0, 0, 0)$ & $1_{c}'$, $1_{abc}'$ \\
      & $\left( \bm{a}, \bm{b}, -\bm{a}+\bm{c}; 0, \frac{1}{4}, 0 \right)$ & $1_{ac}'$, $1_{bc}'$ \\
    9.41
      & $(\bm{a}, \bm{b}, \bm{c}; 0, 0, 0)$ & $1_{a}'$, $1_{b}'$ \\
    \hline
  \end{tabular}
\end{table}

\begin{table}
  \caption{Transformations between type-IV MSG $\mathcal{M}_{\mathrm{BNS}}$ and conjugated MSGs, where their XSGs are identical to a space group with type No. 15 in the ITA standard setting.}
  \label{tab:type4_conjugators_15}
  \begin{tabular}{c|cc}
    \hline
    BNS number & Transformation $(\bm{P}, \bm{p})$ & Anti-translations in $(\bm{P}, \bm{p})^{-1}\mathcal{M}_{\mathrm{BNS}}(\bm{P}, \bm{p})$\\ \hline
    15.90
      & $(\bm{a}, \bm{b}, \bm{c}; 0, 0, 0)$ & $1_{c}'$, $1_{abc}'$ \\
      & $\left( \bm{a}, \bm{b}, -\bm{a}+\bm{c}; \frac{1}{4}, \frac{1}{4}, 0 \right)$ & $1_{ac}'$, $1_{bc}'$ \\
    15.91
      & $(\bm{a}, \bm{b}, \bm{c}; 0, 0, 0)$ & $1_{a}'$, $1_{b}'$ \\
    \hline
  \end{tabular}
\end{table}



\end{document}